\documentclass[aps,prl,twocolumn,amsmath,amssymb,superscriptaddress,nofootinbib,showpacs]{revtex4-1}
\usepackage{aas_macros}
\usepackage{graphicx}
\usepackage[colorlinks=true]{hyperref}

\begin{document}

\title{Testing General Relativity with stellar orbits around the supermassive black hole in our Galactic center}

\author{A. Hees}
\email{ahees@astro.ucla.edu}
\affiliation{Department of Physics and Astronomy, University of California, Los Angeles, CA 90095, USA}

\author{T.~Do}
\affiliation{Department of Physics and Astronomy, University of California, Los Angeles, CA 90095, USA}

\author{A.~M.~Ghez}
\email{ghez@astro.ucla.edu}
\affiliation{Department of Physics and Astronomy, University of California, Los Angeles, CA 90095, USA}

\author{G.~D. Martinez}
\affiliation{Department of Physics and Astronomy, University of California, Los Angeles, CA 90095, USA}

\author{S.~Naoz}
\affiliation{Department of Physics and Astronomy, University of California, Los Angeles, CA 90095, USA}

\author{E.~E.~Becklin}
\affiliation{Department of Physics and Astronomy, University of California, Los Angeles, CA 90095, USA}

\author{A.~Boehle}
\affiliation{Department of Physics and Astronomy, University of California, Los Angeles, CA 90095, USA}

\author{S.~Chappell}
\affiliation{Department of Physics and Astronomy, University of California, Los Angeles, CA 90095, USA}

\author{D.~Chu}
\affiliation{Department of Physics and Astronomy, University of California, Los Angeles, CA 90095, USA}

\author{A.~Dehghanfar}
\affiliation{Department of Physics and Astronomy, University of California, Los Angeles, CA 90095, USA}

\author{K.~Kosmo}
\affiliation{Department of Physics and Astronomy, University of California, Los Angeles, CA 90095, USA}

\author{J.~R.~Lu}
\affiliation{Astronomy Department, University of California, Berkeley, CA 94720, USA}

\author{K.~Matthews}
\affiliation{Division of Physics, Mathematics, and Astronomy, California Institute of Technology, MC 301-17, Pasadena, CA 91125, USA}
	
\author{M.~R.~Morris}
\affiliation{Department of Physics and Astronomy, University of California, Los Angeles, CA 90095, USA}

\author{S.~Sakai}
\affiliation{Department of Physics and Astronomy, University of California, Los Angeles, CA 90095, USA}

\author{R.~Sch\"odel}
\affiliation{Instituto de Astrof\'isica de Andaluc\'ia (CSIC), Glorieta de la Astronom\'ia S/N, 18008 Granada, Spain}

\author{G.~Witzel}
\affiliation{Department of Physics and Astronomy, University of California, Los Angeles, CA 90095, USA}

\date{\today}

\begin{abstract}
	In this Letter, we demonstrate that short-period stars orbiting around the supermassive black hole in our Galactic Center can successfully be used to probe the gravitational theory in a strong regime.  We use 19 years of observations of the two best measured short-period stars orbiting our Galactic Center to constrain a hypothetical fifth force that arises in various scenarios motivated by the development of a unification theory or in some models of dark matter and dark energy.  No deviation from General Relativity is reported and the fifth force strength is restricted to an upper 95\% confidence limit of $\left|\alpha\right| < 0.016$ at a length scale of $\lambda=$ 150 astronomical units. We also derive a 95\% confidence upper limit on a linear drift of the argument of periastron of the short-period star S0-2 of $\left|\dot \omega_\textrm{S0-2} \right|< 1.6 \times 10^{-3}$ rad/yr, which can be used to constrain various gravitational and astrophysical theories. This analysis provides the first fully self-consistent test of the gravitational theory using orbital dynamic in a strong gravitational regime, that of a supermassive black hole. A sensitivity analysis for future measurements is also presented. 

\end{abstract}

\maketitle

\begin{figure}[htbp]
\begin{center}
\includegraphics[width=0.97\linewidth]{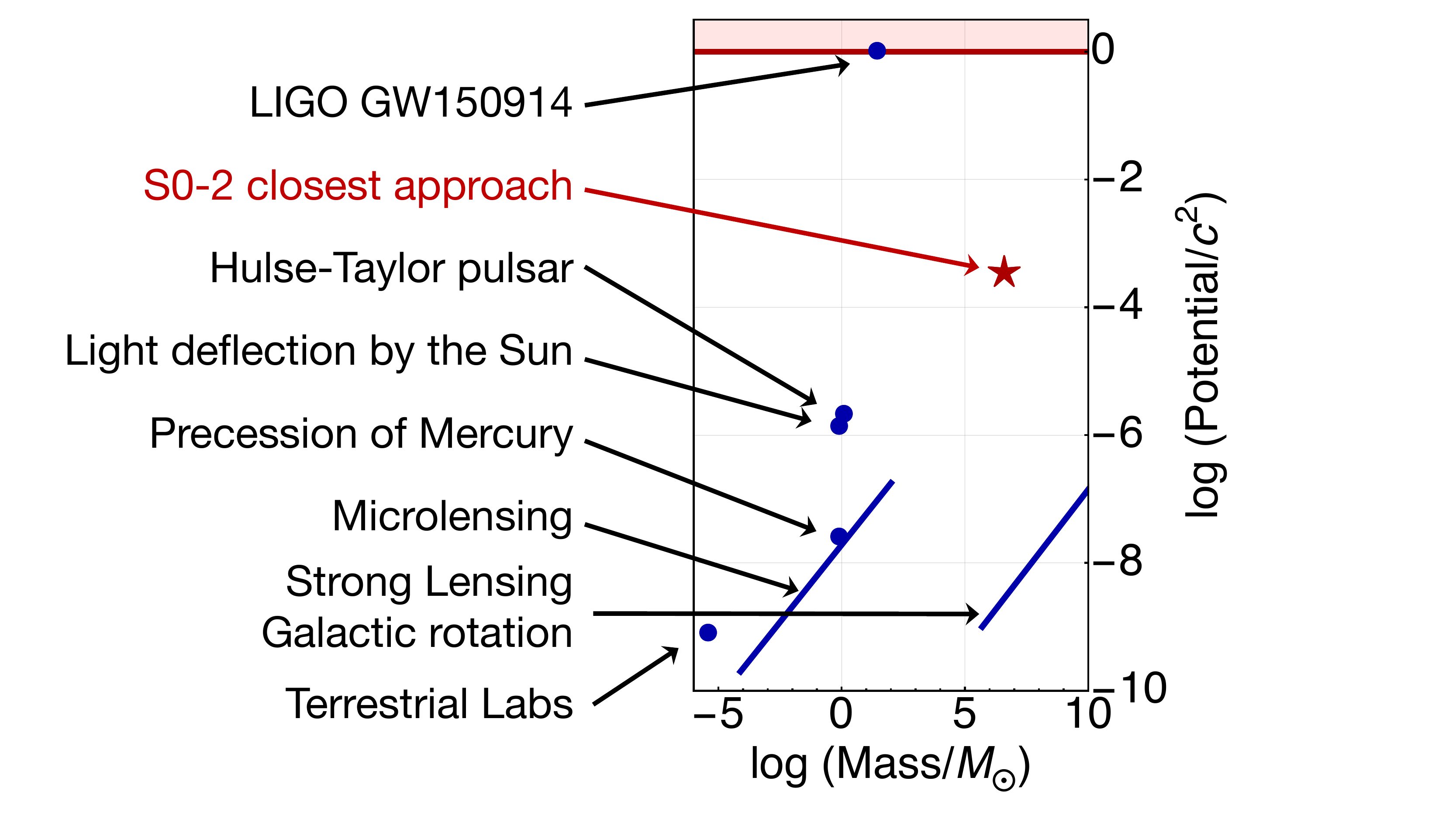}
\end{center}
\caption{The gravitational potential probed by different tests of gravitation  against the mass of the central body that generates gravity in these tests. Short-period stars, such as S0-2, around our Galactic Center explore a new region in this parameter space. The figure is inspired by Ref. \cite{psaltis:2004fk}.}
\label{fig:intro}
\end{figure}

The development of a quantum theory of gravitation or of a unification theory generically predicts deviations from General Relativity (GR). In addition, observations requiring the introduction of dark matter and dark energy also challenge GR and the standard model of particle physics \cite{debono:2016zr} and are sometimes interpreted as a modification of gravitational theory (see, e.g., Refs. \cite{clifton:2012fk,famaey:2012fk}). It is thus important to test the gravitational interaction with different types of observations \cite{berti:2015ve}.   While GR is thoroughly tested  in the Solar System (see, e.g., Refs. \cite{will:1993fk,will:2014la,turyshev:2009qv,tasson:2016fk,*hees:2016aa}) and with binary pulsars (see, e.g., Ref. \cite{stairs:2003aa,*kramer:2016aa}), observations of short-period stars orbiting the supermassive black hole (SMBH) at the center of our Galaxy allow one to probe gravity in a  strong field regime unexplored so far, as shown in Fig.~\ref{fig:intro} (see also Refs. \cite{psaltis:2008uq,johannsen:2016jh,*johannsen:2016sh}). In this Letter, we report two results: (i) a search for a fifth force around our Galactic Center and (ii) a constraint on the advance of the periastron of the short-period star S0-2 that can be used to constrain various gravitational and astrophysical theories in our Galactic Center. This analysis provides the first fully self-consistent test of the gravitational theory using orbital dynamic in a strong gravitational regime, around a SMBH. The constraints presented in this Letter, resulting from 20 years of observations, are therefore highly complementary with Solar System or binary pulsar tests of gravitation and open a new window to study gravitation. 

One phenomenological framework widely used to search for deviations from GR is the fifth force formalism \cite{fischbach:1986uq,talmadge:1988uq,fischbach:1992fk,fischbach:1999ly,adelberger:2003uq,adelberger:2009fk}, which considers deviations from Newtonian gravity in which the gravitational potential takes the form of a Yukawa potential
\begin{equation}\label{eq:pot}
	U=\frac{GM}{r}\left[1 + \alpha e^{-r/\lambda}\right] \, ,
\end{equation}
with $G$ the Newton's constant, $M$ the mass of the central body, and $r$ the distance to the central mass. This potential is characterized by two parameters: a length $\lambda$ and a strength of interaction  $\alpha$.  A Yukawa potential appears in several theoretical scenarios, such as:  unification theories that predict new fundamental interactions with a massive gauge boson \cite{fujii:1971fk,*taylor:1988pi,*kaplan:2000yu,*krause:2001eu}  ($\lambda$ is then related to the mass of the gauge boson through	$m_g=\hbar/c\lambda$, with $c$ being the speed of light in a vacuum and $\hbar$ the reduced Planck constant)~\cite{fischbach:1992fk}, higher dimensional theories (e.g., Ref. \cite{bars:1986qe,*floratos:1999uq,*kehagias:2000kx,*hoyle:2001fk,*goldberger:1999eu} and the references therein), in braneworld scenarios \cite{arkani-hamed:1998vn,*arkani-hamed:1999ij,*dvali:2000qf,*kogan:2001xy,*antoniadis:2003vn}, theories in which supersymmetry breaking originates at low energy (e.g., Ref. \cite{antoniadis:1998qd,*dimopoulos:1996zr}), certain models of Dark Matter (e.g., Ref. \cite{frieman:1991fj,*gradwohl:1992yq,*stubbs:1993uq,*carroll:2009fh,*carroll:2010xw}),  massive Brans-Dicke theories (e.g., Ref. \cite{perivolaropoulos:2010jk,*alsing:2012qy}), certain scalar-tensor-vector models of gravity \cite{moffat:2006lp}, $f(R)$ gravity \cite{capozziello:2015fk}, etc. Moreover, a massive graviton would also lead to a specific case of Yukawa potential characterized by $\alpha=1$ \cite{zakharov:1970fk,*will:1998rz,*visser:1998fk,*damour:2003ys,*babichev:2010vn,hinterbichler:2012nr}. 

The fifth force phenomenology has motivated many experimental searches at a wide variety of scales:  in the lab \cite{spero:1980ek,*hoskins:1985vn,*mio:1987eu,*moody:1993wd,*mohideen:1998wq,*hoyle:2001fk,*long:2003nx,*chiaverini:2003oq,*hoyle:2004kx,*smullin:2005kl,*decca:2005tg,*kapner:2007fk,*decca:2007hc,*tu:2007cr,*adelberger:2007vn,*geraci:2008dq,*masuda:2009yg,*sushkov:2011ys,*yang:2012lq,*chen:2016kk} (see Refs. \cite{adelberger:2009fk,murata:2015uq} for extended reviews), around Earth  \cite{fischbach:1999ly,fischbach:1999ly,iorio:2007wd,*iorio:2012pi,lucchesi:2014ph,*peron:2014wf}, with lunar laser ranging (LLR) \cite{fischbach:1999ly,muller:2005fe,iorio:2012pi} and with planetary motion  \cite{talmadge:1988uq,fischbach:1999ly,konopliv:2011dq,hees:2014jk,*li:2014mz}. All of the current constraints on a fifth force have been obtained with experiments performed in the gravitational field generated by a  weakly gravitationally interacting body (a test mass in the lab, around the Earth or around the Sun) and in a weak gravitational potential (see Fig.~\ref{fig:intro}). 

Constraints on the fifth force in a much stronger and unexplored gravitational regime can be derived using short-period stars around the $4\times 10^6 M_\odot$ SMBH at the Galactic Center \cite{borka:2013nx,*zakharov:2016xy} as depicted on Fig.~\ref{fig:intro}. The motion of short-period stars orbiting around our Galactic Center, Sagittarius A* (Sgr A*), has been monitored for more than 20 years by two experiments, one carried out at the Keck Observatory \cite{ghez:1998ve,ghez:2000rt,ghez:2003qv,ghez:2005dq,ghez:2005kx,ghez:2008bs,meyer:2012qf,boehle:2016wu} and the other with the New Technology Telescope (NTT) and with the Very Large Telescope (VLT) \cite{genzel:1997zr,eckart:1997ys,schodel:2002bh,eckart:2002qf,eisenhauer:2003ty,eisenhauer:2005dz,gillessen:2009cr,gillessen:2009jk,gillessen:2017aa}. These observations have been the source of many discoveries, starting with that of a supermassive black hole  at the center of our Galaxy \cite{genzel:1997zr,ghez:1998ve}. They also have been extremely powerful for improving our understanding of stellar evolution in a galactic nuclear cluster (see Ref. \cite{alexander:2005ul} for a review) and have been used to determine our distance to the Galactic Center with 2 \% relative accuracy \cite{boehle:2016wu,gillessen:2017aa}. In addition, many theorists have anticipated the possibility of measuring relativistic effects and probing the gravitational theory in an unexplored regime \cite{jaroszynski:1998xp,*fragile:2000jx,*rubilar:2001dq,*weinberg:2005cr,*zucker:2006fk,*kraniotis:2007bx,*will:2008fk,*merritt:2010ud,*angelil:2010qd,*angelil:2010to,*iorio:2011rc,*angelil:2011lq,*sadeghian:2011db,*angelil:2014fv,*zhang:2015yu,*borka:2012nr,*capozziello:2014ai,*zakharov:2014qf,*borka:2016rm} (see also the reviews \cite{alexander:2005ul,psaltis:2008uq,johannsen:2016jh,*johannsen:2016sh}).

In this Letter, we search for a fifth force using 19 years of Keck observations of two short-period stars that have been observed throughout their entire orbit: S0-2 (period $P=15.92$ yr and eccentricity $e=0.89$) and S0-38 ($P=19.2$ yr and $e=0.81$) \cite{boehle:2016wu}. We use only the two short-period stars that have a full orbital phase coverage since stars with low phase coverage produce biases in orbital fits \cite{lucy:2014fk}. The data set used is identical to that which is fully described in Ref. \cite{boehle:2016wu}. It includes three types of observations that will be briefly summarized here: (i) speckle imaging data, (ii) adaptive optics (AO) imaging data, and (iii) spectroscopic data. All of the imaging data used come only from the Keck Observatory since there is insufficient information in the public domain to treat other astrometric data in a consistent way.

The speckle data set used for this study provides astrometric diffraction-limited measurements ($\lambda_0=2.21\mu$m, $\Delta \lambda=0.43\mu$m)  of the central 5" $\times$ 5" of the Galactic Center for 27 epochs between 1995 and 2005 and is presented in detail in Refs. \cite{ghez:1998ve,ghez:2000rt,ghez:2005dq,lu:2005vn}. For each epoch of observation, a large number of frames was obtained using short exposure times and was combined using a reconstruction technique called speckle holography \cite{schodel:2013ys}. The positions and fluxes of stars are determined by fitting the point-spread function using the program \textit{StarFinder} \cite{diolaiti:2000zr}. The typical uncertainty of the astrometric positions with the speckle data is on the order of 1.4 milliarcsecond (mas) for S0-2 and  12 mas for S0-38.

The AO data set provides high-resolution images ($\lambda_0=2.12\mu$m, $\Delta \lambda=0.35\mu$m) of the central 10" $\times$ 10" of our galaxy for 23 epochs of observation between 2005 and 2013 \cite{ghez:2005kx,ghez:2008bs,lu:2009qy,yelda:2010fj,yelda:2014yq,boehle:2016wu}. The laser guide-star adaptive optics \cite{wizinowich:2006qf,van-dam:2006xy} corrects instantaneously for most atmospheric aberrations. AO allows for much more efficient observations at the diffraction limit, resulting in measurements with a signal to noise ratio one order of magnitude better than with the speckle observations. With AO observations, the typical uncertainty of the astrometric position is on the order of 0.16 mas for S0-2 and  2 mas for S0-38.

In addition to the central 10" field, we also use six epochs of observations between 2006 and 2013 designed to measure the position of a set of seven SiO maser stars within a 25" field mosaic frame. We tie the infrared measurements of these maser stars to radio astrometric observations \cite{reid:2007nr} to construct an absolute reference frame with Sgr A* at rest \cite{yelda:2010fj,yelda:2014yq}. This is used in order to combine all the speckle holography and AO observations to a common absolute reference frame \cite{boehle:2016wu}.

The third set of data consists of 47 epochs of  spectroscopic observations between 2000 and 2013 \cite{ghez:2003qv,ghez:2008bs,boehle:2016wu}. The procedure used to extract spectra is fully described in Refs. \cite{ghez:2008bs,do:2009ly,do:2013ty}.  The radial velocity (RV) of the stars  is measured using a Gaussian fit to the Br-gamma hydrogen line at 2.1661$\mu$m from the hot atmosphere of S0-2 \cite{ghez:2008bs,do:2013ty} while a cross-correlation method is used for late-type stars like S0-38 \cite{boehle:2016wu}. These RVs are then transformed to the local standard of rest using the ``rvcorrect'' task from the Image Reduction and Analysis Facility (IRAF). In this analysis, we also use RVs measured at the VLT \cite{gillessen:2009cr} similarly. The typical RV uncertainty is 30~km/s for S0-2 and 50~km/s for S0-38.

In total, we use 38 astrometric observations and 47 spectroscopic measurements of S0-2 and  33 astrometric observations and 2 spectroscopic measurements of S0-38 \cite{boehle:2016wu}. Our orbital fits are performed using Bayesian inference with a MultiNest sampler \cite{feroz:2008qf,*feroz:2009xy} using a code that was originally developed in Ref. \cite{ghez:2000rt} and which has been modified to become more flexible over time \cite{ghez:2003qv,ghez:2005dq,ghez:2008bs,meyer:2012qf}.  We extended this code to include the fifth force. The model used for our orbital fits includes the fifth force interaction due to the central SMBH, the R\o mer time delay, the relativistic redshift and the perturbation due to an extended mass. The extended mass density profile is given by a power law \cite{bahcall:1976fk} such that the extended mass enclosed within the radius $r$ is given by
\begin{equation}
	M_\textrm{ext}(<r)=M_\textrm{ext}(<r_0)\left(\frac{r}{r_0}\right)^{3-\gamma}\, .
\end{equation}
We set the outer radius cutoff $r_0$ to 0.011 pc such that it encloses S0-2 and S0-38 at apoapse. In total, our model includes 21 parameters consisting of the six orbital parameters for  each of the two stars:  $P$,  $e$, the time of closest approach $T_0$, the argument of periastron $\omega$, the inclination and the longitude of the ascending node and nine global parameters: the SMBH gravitational parameter $GM$, the strength of the fifth force $\alpha$, the amount of extended mass, $M_\textrm{ext}(<r_0)$, the distance to our Galactic Center, $R_0$, and the positions ($x_0$ and $y_0$) and velocities ($v_{x_0}$,  $v_{y_0}$, $v_{z_0}$) of the SMBH. The SMBH positions and velocities are important in order to take into account imperfections in the construction of the reference frame. The observations are assumed to be independent and normally distributed and we use a Gaussian likelihood analytically marginalized with respect to the SMBH positions and velocities.  We use flat priors for all of the parameters except for the extended mass $M_\textrm{ext}(<r_0)$. The limits for our flat priors have been chosen wide enough to not impact our result (see also Ref. \cite{boehle:2016wu}). Regarding the extended mass, we use an exponential prior characterized by a standard deviation of $\sigma_{M_\textrm{ext}(<r_0)}=100 M_\odot$. This prior is motivated by observations of the stellar cusp \cite{schodel:2009fk,do:2013xy,chappell:2016yu,schodel:2017cr}. The extended mass power-law slope $\gamma$ is fixed to 0.5. We have checked that our results are not sensitive to the actual value of $\sigma_{M_\textrm{ext}(<r_0)}$ and $\gamma$. In particular, $\sigma_{M_\textrm{ext}(<r_0)}$ can be increased by two orders of magnitude without impacting our results. 

From the sampling of the  posterior probability distribution function of $\alpha$,  we determine a statistical 95\% confidence upper limit on the absolute value of $\alpha$. It was shown in Ref. \cite{boehle:2016wu} that our orbital fits suffer from systematic effects related to the construction of the absolute reference frame. In order to assess these systematics, we used a Jackknife resampling method \cite{lupton:1993aa,*gottlieb:2003aa}. We used the seven different reference frames created in Ref. \cite{boehle:2016wu} in which each one has one SiO maser excluded. The results of the orbital fits performed using these seven subset reference frames are then used in order to infer a systematic uncertainty (see Appendix C of Ref. \cite{boehle:2016wu} for more details about this procedure). This inferred systematic uncertainty is then added in quadrature to the statistical uncertainty derived from the orbital fit.  The values of our analysis before and after the Jackknife procedure can be found in Table~1 from the Supplemental material.

\begin{figure*}[htb]
\begin{center}
\includegraphics[width=.8\linewidth]{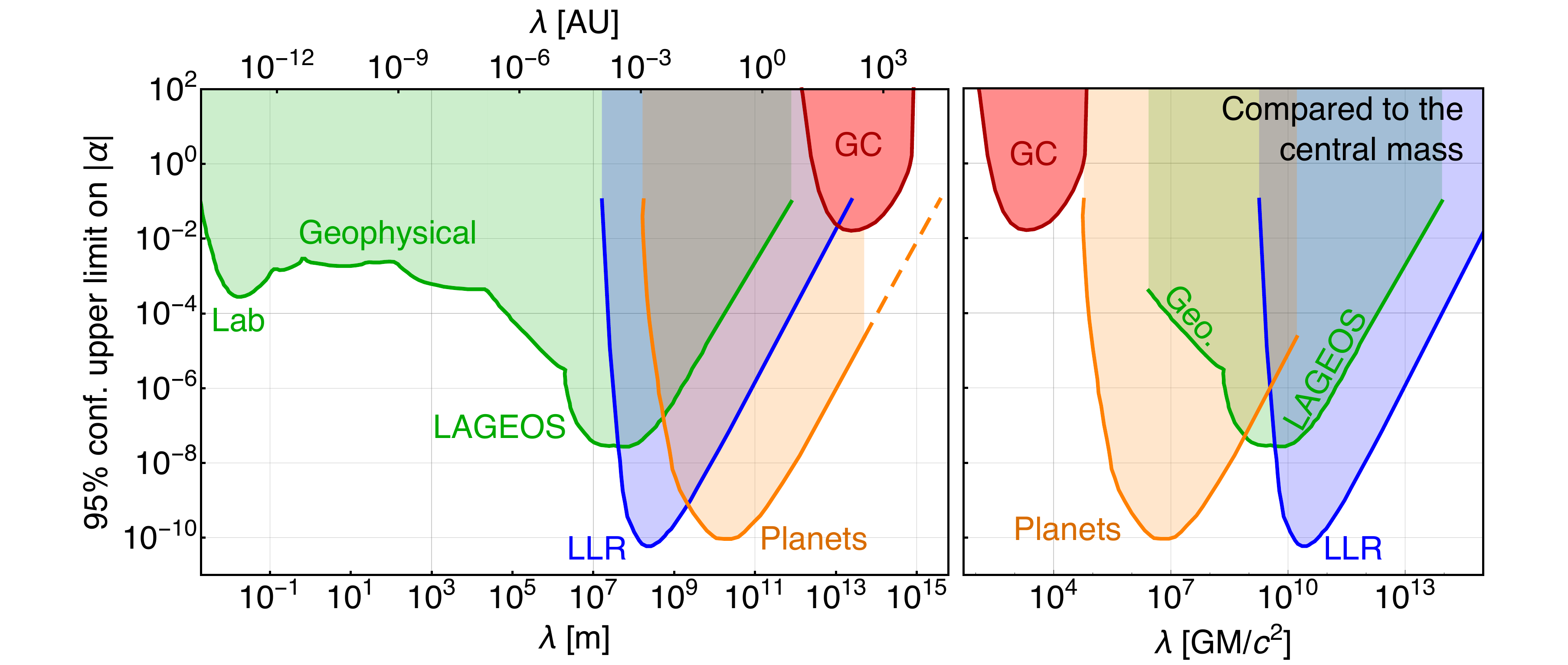}
\end{center}
\caption{95\% confidence upper limits on $\left|\alpha\right|$ as a function of $\lambda$. The shaded regions are excluded by various experiments. Our analysis is represented by the red shaded area (GC) while the other curves are from Fig.~31 of Ref. \cite{konopliv:2011dq}. The dashed curve is a reasonable extrapolation based on Solar System results from Ref. \cite{konopliv:2011dq}. Left panel: the horizontal axis is the fifth force length scale $\lambda$ in meters or in A.U. Right panel: the horizontal axis is the length scale $\lambda$ expressed in term of the gravitational radius of the central mass that generates gravitation in the different experiments. }
\label{fig:all}
\end{figure*}

Our results show that the value $\alpha=0$ is always within the 68\% confidence interval, meaning that no significant deviation from Newtonian gravity is found.  The red curve in Fig.~\ref{fig:all} shows our  95\% confidence upper limit on $\left|\alpha\right|$. Our best constraint is at the level of $\lambda\sim 150$ A.U. which corresponds roughly to the S0-2 distance at periapse. For this value of $\lambda$, our data set gives a 95\% confidence upper limit of $\left|\alpha\right|<0.016$. For higher values of $\lambda$, the upper limit on $\left|\alpha\right|$ evolves proportionally to $\lambda^2$ (similarly to the curves obtained by LLR and planetary ephemerides; see Fig.~\ref{fig:all}) up to when it reaches $\left|\alpha \right|\sim 1$, where it diverges at $\lambda\sim 6000$ A.U. We note that the limit on $\alpha$ is primarily driven by S0-2 in the median range of $\lambda$ and S0-38 helps for small and large $\lambda$'s.

As shown in Fig.~\ref{fig:intro}, the constraints obtained in this work probe a new part of the parameter space and are complementary to Solar System measurements. Specifically, short-period stars are probing space-time in a higher potential and around a central body much more massive than in the other experiments. This is highlighted in the right panel of Fig.~\ref{fig:all}, where $\lambda$ is expressed in terms of the gravitational radius of the central body. Furthermore, short-period stars  probe the space-time around a SMBH, which is conceptually different from Solar System tests where the space-time curvature is generated by weakly gravitating bodies. In particular, some nonperturbative effects may arise around strongly gravitating bodies (see, e.g., Ref. \cite{damour:1993vn}). In addition, in models of gravity exhibiting screening mechanisms, deviations from GR may be screened in the Solar System (see, e.g., Ref. \cite{khoury:2004fk,*khoury:2004uq,*khoury:2010zr,*vainshtein:1972ve,*hees:2012kx,*hinterbichler:2010fk,*hinterbichler:2011uq}). In this context, searches for alternative theories of gravitation in other environments are important.

A specific theoretical model covered by the fifth force framework is a massive graviton. In that context, we found a 90\% confidence limit $\lambda>5000$ A.U. for $\alpha=1$, which can be interpreted as a lower limit on the graviton's Compton wavelength $\lambda_g > 7.5\times 10^{11}$ km or,  equivalently, as an upper bound on the graviton's mass  $m_g < 1.6 \times 10^{-21}$ eV/$c^2$ (see also Ref. \cite{zakharov:2016xy}). This constraint is one order of magnitude less stringent than the recent bound obtained by LIGO \cite{abbott:2016ys,*abbott:2016aa}, which, nevertheless, does not apply for all models predicting a fifth force.

From an empirical perspective, one of the effects produced by a fifth force is  a secular drift of the argument of periastron $\omega$ \cite{burgess:1988fk,iorio:2007wd,iorio:2012pi}. Several theoretical scenarios predict such an effect, which can be constrained by observations. We produced a new orbital fit using a model that includes seven global parameters (the SMBH $GM$, $R_0$ and the positions and velocities of the SMBH) and seven orbital parameters for each star, with the additional parameter being a linear drift of the argument of the periastron $\dot \omega$. As a result of our fit including the Jackknife analysis, we obtained an upper confidence limit on a linear drift of the argument of periastron for S0-2 given by
\begin{align}\label{eq:limit}
	\left|\dot\omega_\textrm{S0-2}\right| &<  1.7 \times 10^{-3} \textrm{ rad/yr} \quad \textrm{at 95 \% C.L.} \, .
\end{align}
This limit is currently one order of magnitude larger than the relativistic advance of the periastron  $\dot \omega_\textrm{GR}=6\pi GM / \left[P c^2 a(1-e^2)\right]=1.6 \times 10^{-4}$ rad/yr for S0-2 (with $a$ being the semimajor axis). Nevertheless, the limit from Eq. (\ref{eq:limit}) can be used to derive a preliminary constraint on various theoretical scenarios (astrophysical or modified gravity) that predict an advance of the periastron for short-period stars in the Galactic Center, like, for example \cite{maeda:2007aa,*gualandris:2010aa,*hackmann:2010aa,*iorio:2011aa,*enolskii:2011aa,*iorio:2012aa,*iorio:2012ab,*iorio:2012ac,*diemer:2013aa,*halilsoy:2013aa,*iorio:2013aa,*capistrano:2014aa,*sotiriou:2014aa,*hu:2014aa,*uniyal:2015aa,*linares:2015aa,*charmousis:2015aa,*bhattacharya:2015aa,*bhattacharya:2017aa,*li:2016aa}

\begin{figure}[htb]
\begin{center}
\includegraphics[width=0.99\linewidth]{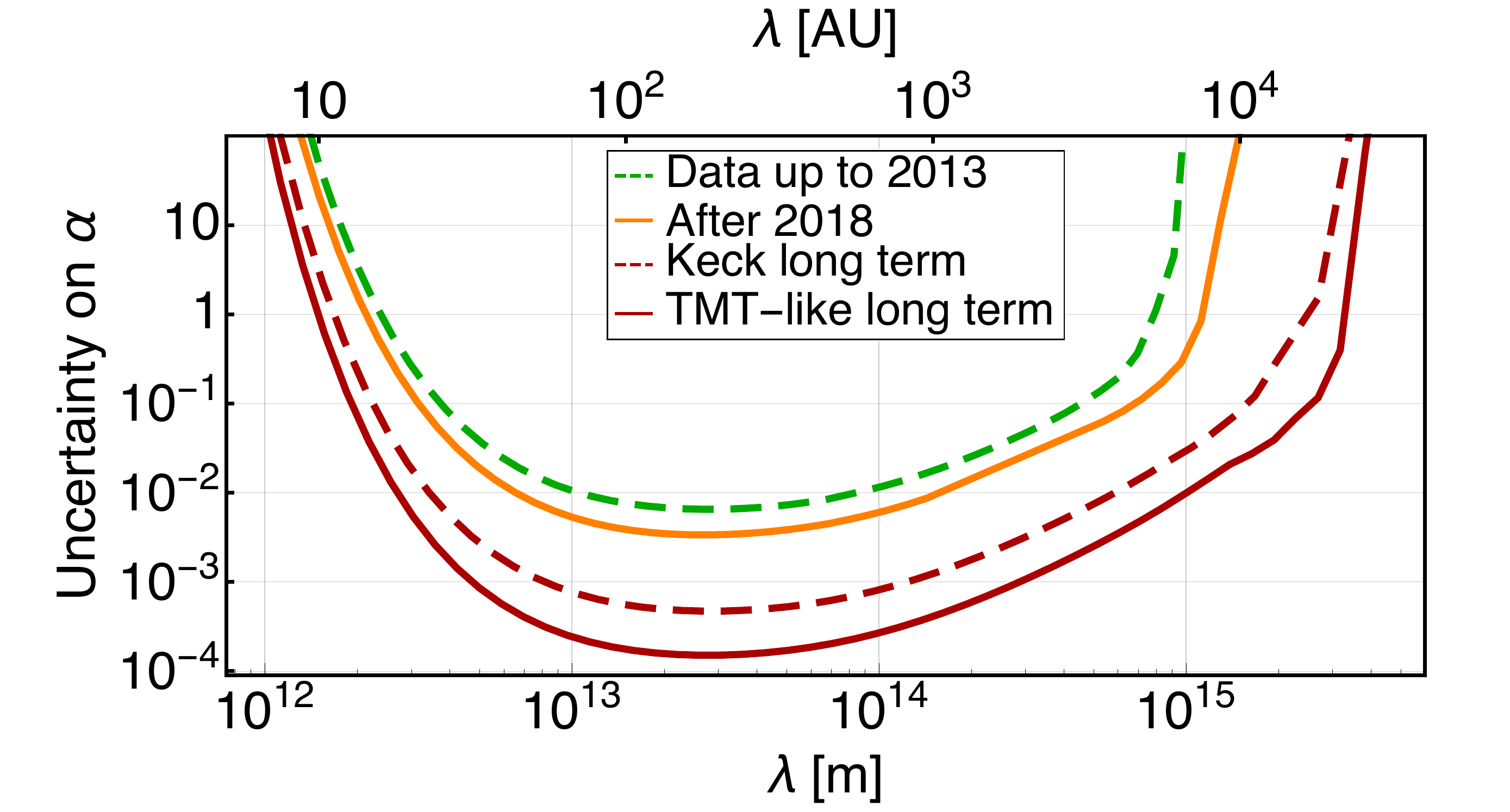}
\end{center}
\caption{Statistical uncertainty on the fifth force strength $\sigma_\alpha$ expected for various observational scenarios: the dashed green (light) line corresponds to the data used in this analysis, the continuous orange (light) line corresponds to data that will be available by the end of 2018. The two red (dark) lines include 16 additional years of observations with 2 astrometric observations and 1 spectroscopic observation per year with the following astrometric/spectroscopic accuracy for an S0-2-like star: current Keck accuracy: 0.5 mas and 30 km/s ; TMT-like improved accuracy : 15 $\mu$as and 5 km/s. }
\label{fig:future}
\end{figure}

Future monitoring of short-period stars will improve the results presented in Fig.~\ref{fig:all}. For example, after the S0-2 closest approach in 2018, our current constraints on $\alpha$ are expected to be improved by a factor of 2 as shown in Fig.~\ref{fig:future}. On a longer term, the next generation of telescopes like the Thirty Meter Telescope (TMT) will significantly improve  the current results. Fig.~\ref{fig:future} shows a sensitivity analysis based on a Fisher matrix approach performed to assess the improvement expected by observations with a TMT-like telescope. We have simulated 16 additional years of data for two scenarios: (i) a scenario where Keck observations are used with an astrometric uncertainty of 0.16 mas, comparable to today's performance and (ii) a scenario with an improved astrometric uncertainty of 0.015 mas which corresponds to a TMT-like scenario. Extending the time baseline by one S0-2 period improves the result by a factor of 13, while an improved accuracy brings an additional improvement of a factor of 5. In addition, the discovery of new stars orbiting closer to the SMBH and unbiased measurements of the known faint short-period star S0-102 ($P=11.5$ yr) \cite{meyer:2012qf} would  improve  this analysis.

In conclusion, we have used 19 years of observations of S0-2 and S0-38 reported in Ref. \cite{boehle:2016wu} to constrain a hypothetical fifth interaction around the SMBH in our Galactic Center. The constraints obtained in our analysis are summarized in Fig.~\ref{fig:all}. Our results are complementary to the ones obtained in the Solar System since they are obtained in a completely different and unexplored strong field regime. We have shown that future observations and especially the next generation of telescopes will improve our results substantially. In addition, we have derived a limit on an hypothetical advance of the periastron of the short-period star S0-2, a constraint that can be used to constrain various astrophysical and fundamental physics scenarios in the Galactic Center.  This analysis shows that we are currently entering an era where astrometric and spectroscopic observations of short-period stars around Sgr A* can be used to probe fundamental physics. This will be reinforced with the detection of the relativistic redshift after the S0-2 closest approach in 2018, as anticipated in Ref. \cite{will:2008fk}. In  the longer term, tests of GR using short-period stars are expected to complement other types of observations that will probe the space-time around the SMBH at the center of our Galaxy, such as, for example, observations made with the Event Horizon Telescope \cite{johannsen:2016kx,*johannsen:2016uq,psaltis:2016jk}.

\par {\it Acknowledgments.} A.H. thanks W. Folkner and P. Wolf for interesting discussions. Support for this work was provided by NSF Grant No. AST-1412615, the Heising-Simon Foundation, the Levine-Leichtman Family Foundation, the Galactic Center Board of Advisors, and Janet Marott for her support of the research on S0-38 through the Galactic Center Stellar Patron Program.

\bibliography{../../../JPL/JPL_byMe/biblio}
\end{document}